\documentclass[preprint,amsmath,amssymb]{revtex4}
\usepackage{graphicx}

\usepackage{amssymb}

\def\a{\mbox{\boldmath $a$}}\def\b{\mbox{\boldmath $b$}}
\def\c{\mbox{\boldmath $c$}}

\def\v{\mbox{\boldmath $v$}}
\def\w{\mbox{\boldmath $w$}}\def\x{\mbox{\boldmath $x$}}
\def\y{\mbox{\boldmath $y$}}\def\z{\mbox{\boldmath $z$}}
\def\A{\mbox{\boldmath $A$}}
\def\C{\mbox{\boldmath $C$}}

\def\I{\mbox{\boldmath $I$}}

\def\O{\mbox{\boldmath $O$}}

\def\S{\mbox{\boldmath $S$}}
\def\U{\mbox{\boldmath $U$}}
\def\X{\mbox{\boldmath $X$}}

\def\0{\mbox{\bf 0}}\def\1{\mbox{\bf 1}}\def\2{\mbox{\bf 2}}
\def\3{\mbox{\bf 3}}\def\4{\mbox{\bf 4}}\def\5{\mbox{\bf 5}}
\def\6{\mbox{\bf 6}}\def\7{\mbox{\bf 7}}\def\8{\mbox{\bf 8}}
\def\9{\mbox{\bf 9}}

\def\OC{\mbox{$\cal{O}$}}

\def\Real{\mbox{$\mathbb R$}}
\def\NaturalNumber{\mbox{$\mathbb N$}}

\def\KK{\mbox{$\mathbb K$}}
\def\SS{\mbox{$\mathbb S$}}
\def\Diag{\mbox{\bf Diag}}
\def\gam{\mbox{$\gamma$}}
\def\Gam{\mbox{$\Gamma$}}

\def\eg{\mbox{\it e.g.}}

\makeindex



\begin{document}

\setcounter{page}{1}  

\title{VARIATIONAL APPROACH FOR THE ELECTRONIC STRUCTURE
CALCULATION ON THE SECOND-ORDER REDUCED DENSITY MATRICES
AND THE $N$-REPRESENTABILITY PROBLEM}

\markboth{M.~Nakata, M.~Fukuda, \& K.~Fujisawa}{Variational approach for the electronic structure calculation}

\author{Maho Nakata} 

\affiliation{Advanced Center for Computing and Communication\\
RIKEN\\
2-1 Hirosawa, Wako-city, Saitama 351-0198 Japan\\
maho@riken.jp}

\author{Mituhiro Fukuda}  

\affiliation{Department of Mathematical and Computing Sciences\\
Tokyo Institute of Technology\\
2-12-1-W8-41 Ookayama, Meguro-ku, Tokyo 152-8552 Japan\\
mituhiro@is.titech.ac.jp} 

\author{Katsuki Fujisawa}

\affiliation{Department of Industrial and Systems Engineering\\
Chuo University\\
1-13-27 Kasuga, Bunkyo-ku, Tokyo 112-8551 Japan\\
fujisawa@indsys.chuo-u.ac.jp} 

\date{20 October 2010, revised 23 June 2011}

\begin{abstract}
The reduced-density-matrix method is a promising candidate for the next generation electronic structure calculation method; it is equivalent to solve the Schr\"odinger equation for the ground state. The number of variables is the same as a four electron system and constant regardless of the electrons in the system. Thus many researchers have been dreaming of this much simpler method for quantum mechanics. In this chapter, we give a overview of the reduced-density-matrix method; details of the theories, methods, history, and some new computational results. Typically, the results are comparable to the CCSD(T) which is a sophisticated traditional approach in quantum chemistry.
\end{abstract}


\maketitle

\section{Introduction}
Chemistry is an important branch of science which treats change of matter. It
explains, for example, why and how a protein works, 
the process $\rm CO_2$ converts to $\rm O_2$, {\it etc.}
The goal is to
predict, understand, and control 
what will happen when we mix some substances. To do that, we usually do
experiments which can be explosive, poisonous, expensive and unstable.
Therefore, it is desirable to do chemistry without experiments. Fortunately,
the basic equation for chemistry is already known
and it is called the Schr\"odinger equation \cite{Dirac29}. 
It is possible to approximately
solve the Schr\"odinger equation using computers with various 
methods and algorithms.
Such branch of chemistry is called quantum chemistry and it is our main 
interest\footnote{also known as theoretical chemistry or computational chemistry.}.

Determining the exact or approximate solution to the Schr\"odinger equation is the fundamental problem in quantum chemistry. This solution is called the {\it wavefunction}, or sometimes referred as {\it electronic structure}. If we know the electronic structure, we can do chemistry. Often such methods are referred as {\it ab initio}  (Latin word which means ``from the beginning'') or {the first principle} method if approximations are not heuristic or do not employ parameters from experiments.

The {\it ground state energy} calculation of a non-relativistic and time-independent, $N$-electron molecular system under the Born-Oppenheimer approximation is the most important problem \cite{SzaboOstlund}. It can be obtained as the lowest eigenvalue $E$ of the electronic Schr\"odinger equation:

\begin{equation}
\label{eq:schrodinger}
H \Psi(\z) = E \Psi(\z), 
\end{equation}
where $H$ is the {\it Schr\"odinger operator} or 
{\it Hamiltonian} defined by

\begin{equation}
\label{eq:molehamiltonian}
H = -\frac{1}{2}\sum_{i=1}^N \nabla^2_i - \sum_{i=1}^N\sum_{A=1}^M
\frac{Z_A}{r_{iA}}+\sum_{i=1}^N\sum_{j>i}^N\frac{1}{r_{ij}},
\end{equation}
in which $Z_A$ is the atomic number of the nucleus $A$,
$r_{iA}$ is the distance between the electron $i$ and nucleus $A$,
and $r_{ij}$ is the distance between two distinct electrons.
The solution of (\ref{eq:schrodinger}), $\Psi(\z)$ 
in $L^2(\KK^N)$, $\KK=\Real^3\times \{-\frac{1}{2},\frac{1}{2}\}$ with
the inner product $\langle\Psi_1(\z),\Psi_2(\z)\rangle =
\int\Psi_1(\z)\Psi_2(\z)d\z, \quad \z=(\x,s)\in
\KK$ ($\int d\z$ includes integration over spin variables), 
is the {\it wavefunction}, and the corresponding eigenvalue $E$,
the {\it total energy} of the system.

Besides, since electrons are fermions, the wavefunction itself is antisymmetric due to Pauli exclusion principle:
\[
\Psi(\z_1,\ldots,\z_i,\ldots,\z_j,\ldots,\z_N) = -\Psi(\z_1,\ldots,\z_j,\ldots,\z_i,\ldots,\z_N).
\]
That is to say, we must solve the Schr\"odinger equation in the antisymmetric subspace of $L^2(\KK^N)$. We denote such space as ${\cal A}L^2(\KK^N)$.

Even on computers, treating the $N$-particle wavefunction is very difficult. Thus, we discretize the Hilbert space ${\cal A}L^2(\KK^N)$ by taking antisymmetric products of the one-particle Hilbert space $L^2(\KK)$, whose complete orthonormal system (CONS) is $\{ \psi_i \}_{i=1}^{\infty}$. Each $\psi_i$ is called 
{\it single-electron wavefunctions} or {\it spin orbitals}
\begin{equation}
\label{eq:basis}
\psi_i: \KK \rightarrow \Real. \quad (i=1,2,\ldots, \infty)
\end{equation}
We can explicitly construct a CONS of ${\cal A}L^2(\KK^N)$ using $\{ \psi_i \}_{i=1}^{\infty}$ by the Slater determinants defined as follows \cite{SzaboOstlund}:
\[
\Psi_I (\z) = \frac{1}{\sqrt{N!}}
\left|
\begin{array}{cccc} 
     \psi_{i_1} (\z_1) & \psi_{i_2} (\z_1) & \cdots & \psi_{i_{N}} (\z_1) \\
     \psi_{i_1} (\z_2) & \psi_{i_2} (\z_2) & \cdots & \psi_{i_{N}} (\z_2) \\
      \vdots & \vdots &  \ddots     & \vdots \\ 
     \psi_{i_1} (\z_{N}) & \psi_{i_2} (\z_N) & \cdots & \psi_{i_{N}} (\z_{N}) \\
\end{array}
\right|.
\]
Here, we used an ordered set of indices $I=\{i_1,\ldots,i_j,\ldots,i_k,\ldots,i_N\} \ \textrm{where} \ i_j < i_k, \ i_j,i_k\in \NaturalNumber$. It is known that $\{\Psi_I\}$ is a CONS of ${\cal A}L^2(\KK^N)$ \cite{Lowdin55}. 

A second approximation to solve the Schr\"odinger equation would be choose carefully $r$
functions from a CONS $\{ \psi_i \}_{i=1}^{\infty}$ by chemical or physical intuition. Then we construct a subspace of ${\cal A}L^2(\KK^N)$ by the Slater determinants considering all possible combinations of $N$ spin orbitals among
$r$ possibilities. In this case, solving the Schr\"odinger equation becomes the eigenvalue problem of the Hamiltonian matrix which now seems to be feasible on computers. Nevertheless, the dimension of the problem becomes
$r!/(N!(r-N)!)$ which obviously is impractical even for small values.
The (approximate) ground state energy obtained by this procedure is considered the reference value, and called
{\it Full Configuration Iteration} (Full CI) (energy).
The mainstream approaches in quantum chemistry can be roughly
interpreted as
linear or nonlinear approximations of this eigenvalue problem, \eg, Hartree-Fock method, second-order perturbation methods, coupled cluster methods, truncated CI methods, {\it etc.}

Our main motivation is employ the second-order reduced density matrix (2-RDM) as the basic variable for quantum mechanics to construct simpler methods. Since only two-body interactions exist in nature, we can calculate all observables using the 2-RDM. Moreover, the number of variables of the 2-RDM is always four, regardless of the number of electrons in the system ($r^4$ when discretized), whereas the wavefunction scales like $N$ ($r!/(N!(r-N)!$ when discretized).

This chapter is organized as follows. In Section~2, we define the
first-order and second-order reduced density matrices, and introduce
the notion of $N$-representability and its conditions. The
reduced-density-matrix method is a viable implementation to approximate
the ground state energy of molecular systems. The reduced-density-matrix
method is formulated as an semidefinite program in Section~3, and
its numerical results using a parallel optimization code is given in
Section~5. In Section~4, we give a brief historical note of this
approach. Finally in Section~6, we give some concluding remarks.

\section{The Reduced-Density-Matrix Method}
\subsection{Pure states and ensemble states}
Most generally, a quantum system containing $N$ particles is described by the density matrix $D$ which was introduced independently by von Neumann, Landau, and Bloch, and is an ensemble average
of wavefunctions,
\[
D (\z_1,\z_2,\ldots,\z_N, \z_1^\prime,\z_2^\prime,\ldots,\z_N^\prime) = \sum_{i} w_i \Psi_i(\z_1,\z_2,\ldots,\z_N)\Psi_i^*(\z_1^\prime,\z_2^\prime,\ldots,\z_N^\prime),
\]
where $w_i \geq 0$, $\sum_{i=1}^\infty w_i = 1$, and $\{\Psi_i \}_{i=1}^{\infty}$ is a CONS of an $N$-particle state.
For a pure state where the system is described by the wavefunction $\Psi$, $D$ can be written by 
\[
D(\z_1,\z_2,\ldots,\z_N, \z_1^\prime,\z_2^\prime,\ldots,\z_N^\prime) = \Psi(\z_1,\z_2,\ldots,\z_N)\Psi^*(\z_1^\prime,\z_2^\prime,\ldots,\z_N^\prime).
\]
This is equivalent to requiring $D$ to be idempotent; $D^2=D$. Hereafter, when we refer to a state, it means be an ensemble if not otherwise specified. We are mainly interested in the pure state but usually we do not care about whether $D$ is a pure or ensemble. This only becomes a problem when the system is degenerated or we consider its subsystems.

\subsection{The first-order and second-order reduced density matrices}
Given an ensemble $D(\cdot)$, the {\it first-order Reduced Density Matrix} (1-RDM) \cite{Dirac29} is defined by
\[
\gamma(\z_1,\z_1^\prime) = N \int D (\z_1,\z_2,\ldots,\z_N, \z_1^\prime,\z_2,\ldots,\z_N)d\z_2d\z_3\cdots d\z_N.
\]
The {\it second-order Reduced Density Matrix} (2-RDM) \cite{Husimi40,Lowdin55,Mayer55} is given by
\[
\Gamma(\z_1,\z_2,\z_1^\prime,\z_2^\prime) = \frac{N (N-1)}{2} \int D (\z_1,\z_2,\ldots,\z_N, \z_1^\prime,\z_2^\prime,\ldots,\z_N)d\z_3\cdots d\z_N,
\]
and higher-order RDMs are defined in an analogous way. The normalization factor for the $p$-th order reduced density
matrix is then $\frac{N!}{p!(N-p)!}$.
The second-quantized versions are defined using a set of creation and annihilation operators $\{a_i, a_i^\dagger\}_{i=1}^{\infty}$, where $a_i^\dagger$ 
creates and $a_i$ annihilates a particle at $\psi_i(\z)$ of $D$ as follows \cite{SzaboOstlund}:
\begin{eqnarray}
\gamma^i_j & = &  \textrm{tr} (a^\dagger_i a_j D)  = \sum_{p} w_p \langle \Psi_p | a^\dagger_i a_j | \Psi_p \rangle, \nonumber \\ \label{eq:1rdm}  
\Gamma^{ij}_{k\ell} & = & \frac{1}{2} \textrm{tr} (a^\dagger_i a^\dagger_j a_{\ell} a_k D) = \frac{1}{2}\sum_p w_p \langle \Psi_p | a^\dagger_i a^\dagger_j  a_{\ell} a_k | \Psi_p \rangle.
\label{eq:2rdm} \nonumber
\end{eqnarray}
The normalization factor for the $p$-th order reduced density matrix is $1/p!$ for this case. 
The equivalence of these two different expressions can be found by
\begin{eqnarray}
\gamma^i_j & = & \int \psi^*_i(\z_1) \gamma(\z_1,\z_1^\prime) \psi_j(\z_1^\prime) d\z_1 d\z_1^\prime, \nonumber \\
\Gamma^{ij}_{k\ell} & = & \int \psi^*_i(\z_1) \psi^*_j(\z_2) \Gamma(\z_1,\z_2,\z_1^\prime,\z_2^\prime) \psi_{\ell}(\z_1^\prime) \psi_k(\z_2^\prime) d\z_1 d\z_2 d\z_1^\prime d\z_2^\prime,  \nonumber
\end{eqnarray}
where we used the single-particle wave function $\{ \psi_i(\z) \}_{i=1}^{\infty}$ of (\ref{eq:basis}).
The following conditions are inherited by these definitions:
\begin{description}
\item {(1)} the 1-RDM and 2-RDM are Hermitian,
\[
\gamma^i_j = ( \gamma^j_i )^* , \quad \Gamma^{ij}_{k\ell} = ( \Gamma^{k\ell}_{ij} )^*,
\]

\item {(2)} the 2-RDM is antisymmetric,
\[
\Gamma^{ij}_{k\ell} = - \Gamma^{ji}_{k\ell} = -\Gamma^{ij}_{\ell k} = \Gamma^{ji}_{\ell k},
\]

\item {(3)} trace conditions are valid,
\[
\sum_{i} \gamma^i_i = N, \quad \sum_{ij}\Gamma^{ij}_{ij} = \frac{N(N-1)}{2},
\]

\item {(4)} a partial trace condition holds between the 1-RDM and 2-RDM,
\[
\frac{N-1}{2}\gamma^i_j = \sum_{k}\Gamma^{ik}_{jk}.
\]
\end{description}
Additionally, we can find more conditions from the symmetry of the system. 
In particular, the spin symmetry of the (ground state) molecular systems is important and formulated as follows:
\begin{description}
\item {(5)} the total spin $S^2$; the 2-RDM should be the eigenstate of spin operator
\[
 {\rm tr } (S^2 \Gamma) = S(S+1),
\]
where $S^2$ is defined as follows:
\begin{eqnarray}
S^2 & = & S_x^2 + S_y^2 + S_z^2 = S_z +S_z^2 + S_-  S_+ \nonumber  \\
&= & \frac{1}{2} \sum_{i} \left ( a_{i\alpha}^\dagger a_{i\alpha}-
a_{i\beta}^\dagger a_{i\beta} \right ) + \frac{1}{4} \left (\sum_{i} a_{i\alpha}^\dagger a_{i\alpha}-   a_{i\beta}^\dagger a_{i\beta}  \right )^2 \nonumber \\
& & + \sum_{ij} a_{i\beta}^\dagger a_{i\alpha} a_{j\alpha}^\dagger a_{j\beta} \nonumber
\end{eqnarray}
The indices $i\alpha$, $j\beta$ means that we choose spin eigenfunctions of the $z$-axis for $\{ \psi_i(\z) \}_{i=1}^{\infty}$, and reorder them so that $i$ means $i$-th spacial function and $\alpha$, $\beta$ denote eigenfunctions of $\alpha$-spin and $\beta$-spin; $\{ \psi_{i\alpha}(\z), \psi_{i\beta}(\z) \}_{i=1}^{\infty}$, respectively.
\item {(6)} The $z$-component of the spin, $S_z$ can be chosen as integer or half integer,
\[
\langle S_z \rangle = \frac{1}{2}\sum_{i} (\gamma^{i\alpha}_{i\alpha} - \gamma^{i\beta}_{i\beta}).
\]
\end{description}
In the subsequent discussion, one will notice that the 1-RDM can be
disregarded throughout. However, we explicitly use it in order to prioritize
the compactness of the notation. 

\subsection{Solving the ground state problem using 1- and 2-RDMs}

The Hamiltonian of the most general form in second-quantization can be written by:
\[
 H = \sum_{ij} v^{i}_{j} a_i a_j^\dagger + \frac{1}{2}\sum_{ijk\ell} w^{ij}_{k\ell} a_i a_j a^\dagger_{\ell} a^\dagger_k,
\]
where $v^{i}_{j}$ and $w^{ij}_{k\ell}$ are one- and two-particle terms which can be calculated from the molecular Hamiltonian (\ref{eq:molehamiltonian}) by the Slater's rule as follows:
\[
 v^{i}_{j} = \int \psi^*_i(\z) \left (-\frac{1}{2} \nabla^2 - \sum_{A=1}^M\frac{Z_A}{r_{A}} \right ) \psi_j(\z)d\z,
\]
\[
 w^{ij}_{k\ell} = \int \psi^*_i(\z_1) \psi^*_j(\z_2) \left ( \frac{1}{|\z_1 - \z_2|}\right ) \psi_{\ell}(\z_1) \psi_k(\z_2)  d\z_1 d\z_2,
\]
where $r_A$ is the distance between an electron and a nucleus.
Then, the total energy $E$ can be expressed using $D$ as follows:
\[
 E = \textrm{tr}(HD).
\]
The ground state energy $E_{\rm min}$ can be calculated minimizing the total energy
over 1- and 2-RDMs.
\begin{eqnarray}
 E_{\rm min} & = & {\rm min} \,\textrm{tr} (H D) \nonumber \\
     & = & {\rm min} \,\textrm{tr} ( \sum_{ij} v^{i}_{j} a_i a_j^\dagger + \frac{1}{2}\sum_{ijk\ell} w^{ij}_{k\ell} a_i a_j a^\dagger_{\ell} a^\dagger_k ) D \label{eq:wfmin} \\
     & = & {\rm min} \sum_{ij} v^{i}_{j} \textrm{tr} ( a_i a_j^\dagger D ) + \sum_{ijk\ell} w^{ij}_{k\ell} \frac{1}{2}\textrm{tr} ( a_i a_j a^\dagger_{\ell} a^\dagger_k D ) \nonumber \\
     & = & {\rm min} \{ \sum_{ij} v^{i}_{j} \gamma^{i}_{j} +  \sum_{ijk\ell} w^{ij}_{k\ell} \Gamma^{ij}_{k\ell} \}. \label{eq:rdmmin}
\end{eqnarray}
It is easy to show that this minimization (\ref{eq:wfmin}) is equivalent to solve the Schr\"odinger equation for the ground state, and often such kind of methods are called variational methods. Moreover, (\ref{eq:rdmmin}) is also equivalent to solve the Schr\"odinger equation for the ground state. Advantage of using 1- and 2-RDMs instead of $D$ is that the number of variables are reduced drastically.

\subsection{The $N$-representability problem and the $N$-representability conditions}

In 1950's, researchers based on the above facts chose the 1- and 2-RDMs as basic variables, and did some variational calculations; according to L\"owdin \cite{Lowdin87}, F. London, J. E. Mayer, A. J. Coleman, P. O. L\"owdin, R. McWeeny, N. A. March, C. A. Coulson and others have attempt to minimize via (\ref{eq:rdmmin}). However, their results were considerably lower than the true energy. 
The reason is that the trial 2-RDMs were not actually derived from an existing
density matrix $D$. We need some more conditions on the trial 2-RDM to ensure that it comes from a true $D$. 
Such formalism of the problem was first described by 
A.~J.~Coleman in 1963 and named {\it $N$-representability problem}, and these
conditions are known to be the {\it $N$-representability conditions} \cite{Coleman63}; Given a trial $p$-th order RDM $\Gamma^{(p)}$, if there exists some wavefunction or ensemble which reduces to the $p$-th order RDM $\Gamma^{(p)}$, 
then this $\Gamma^{(p)}$ is pure or ensemble $N$-representable, respectively. 

\subsection{On the complete $N$-representability conditions}
Given a 1- and 2-RDM, they should satisfy the relations $(1)$ to $(4)$ of
Section~2.2
The 1- and 2-RDM for the ground state should additionally satisfy $(5)$ and $(6)$.
Therefore, these conditions are necessary conditions for the
$N$-representability. Unfortunately, these conditions are not sufficient, thus early attempts failed and obtained very low energy.
The necessary and sufficient condition for the 1-RDM is relatively easy \cite{Kuhn60,Coleman63}. However, the complete (sufficient) $N$-representability of the 2-RDM is very complicated in general. Garrod and Percus \cite{GarrodPercus64} showed that the 2-RDM $\Gamma$ is ensemble $N$-representable
if and only if
\[
\sum_{ijk\ell} {(H^\nu)}^{ij}_{k\ell} \Gamma^{ij}_{k\ell} \geq {E^\nu_{\rm min}}, \label{eq:compnrep}
\]
where $H^\nu$ is every possible Hamiltonian and $E^\nu_{\rm min}$ is the ground state energy corresponds to $H^\nu$. Thus, the ensemble $N$-representable set ${\cal E}^N$ can be defined by:
\[
{\cal E}^N = \{ \Gamma \ | \ \textrm{tr}(H^\nu\Gamma) \geq E^\nu_{\rm min}, \textrm{for all possible $H^\nu$ and $E^\nu_{\rm min}$}\}.
\]
This result is theoretical and very important, but totally not practical since if one wants to calculate the {\it exact} ground state energy of a Hamiltonian, then he/she must know the exact ground state energy of the system beforehand. This is a tautology! After that, many researchers seek the complete $N$-representability condition, and did not succeed. A meaningful result from complexity theory was obtained by Liu {\it et al}. \cite{Liu2007} in 2007. They showed that the computational complexity of the $N$-representability problem is QMA-complete, which is the quantum generalization of NP-completeness. Thus it is almost hopeless to find an efficient algorithm to decide whether a given 2-RDM is $N$-representable or not. We can consider a more physical example: the ground state problem of the spin-glass Hamiltonian is known to be a very hard problem, and equivalent to solve the max-cut problem or the traveling sales person problem, which in turn are known to be NP-hard \cite{Barahona88}. If the complete $N$-representability conditions were easy to handle, we could solve such difficult problems in computer science as well. Currently we do not know how to solve these problems efficiently. We just want to stress, the complete $N$-representability is a really hard problem. 


\subsection{Formulating as a variational problem, and its
geometrical representation}
The problem we want to solve can be formulated using the 1- and 2-RDMs as basic variables,
\begin{eqnarray}
 E_{\rm min} & = & {\rm min} \ \textrm{tr} (H D) \nonumber \\
     & = & \underset{\Gamma \in {\cal E}^N}{\rm min} \{ \sum_{ij} v^{i}_{j} \gamma^{i}_{j} +  \sum_{ijk\ell} w^{ij}_{k\ell} \Gamma^{ij}_{k\ell} \}. \nonumber  
\end{eqnarray}

\begin{figure}
\begin{center}
\includegraphics[scale=0.4, bb=100 200 455 635]{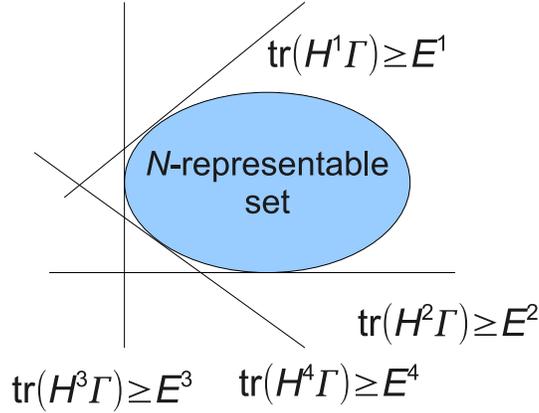}
\caption{Schematic representation of the $N$-representable set; a Hamiltonian and its ground state energy serves as a characterization of the set.}
\label{fig:nrep}
\end{center}
\end{figure}

${\cal E}^N$ is known to be a compact convex set. Besides, all possible
Hamiltonians and the corresponding ground state energies serves as a characterization of this convex set. To be more precise, a 2-RDM corresponding to the ground state of an $N$-particle Hamiltonian is a surface point, and any surface point of ${\cal E}^N$ corresponds to the ground state of some Hamiltonian \cite{Rosina}. 
The compact and convex set of the $N$-representable set is represented as an ellipse (we do not show, but there are also cusp points as well) in Fig. \ref{fig:nrep}. The Hamiltonians $H^1$, $H^2$, $H^3$, and $H^4$, and their ground state energies $E^1$, $E^2$, $E^3$, and $E^4$ serves as $N$-representability conditions, respectively. 

\subsection{Some known necessary $N$-representability conditions}
We should not be demotivated by the facts of the previous subsection. Understanding the chemical and/or physical meaning of the necessary $N$-representability conditions is much more important. Mathematical theorems do not tell about chemistry or physics. In practice, $N$-representability conditions for the molecular systems might not be so difficult. 

We seek for chemically and/or physically meaningful necessary $N$-representability conditions on the 1- and 2-RDMs.
The necessary and sufficient conditions for an ensemble $N$-representability for the 1-RDM is characterized by its eigenvalues lying between 0 and 1
\cite{Kuhn60,Coleman63}. 
For the pure state, it is more complicated \cite{Klyachko08}. Coleman introduced two necessary conditions called the $P$ and $Q$ conditions \cite{Coleman63}. These conditions require the positive semidefiniteness of the $P$-matrix ($\Gamma$), and the $Q$-matrix
\begin{eqnarray}
 P^{ij}_{k\ell} & = & \textrm{tr}(a^\dagger_i a^\dagger_j a_{\ell} a_k D ) \succeq 0, \label{p} \\ 
 Q^{ij}_{k\ell} & = & \textrm{tr}(a_i a_j a^\dagger_{\ell} a^\dagger_k D ) \succeq 0. \label{q}
\end{eqnarray}
Another important necessary condition is called the $G$ condition \cite{GarrodPercus64}, which also require positive semidefiniteness of the $G$-matrix defined as follows
\begin{eqnarray}
 G^{ij}_{k\ell} & = & \textrm{tr} (a^\dagger_i a_j a^\dagger_{\ell} a_k D ) \succeq 0. \label{g}
\end{eqnarray}
In the original paper by Garrod-Percus, the definition of the $G$-matrix is non-linear:
\begin{eqnarray}
 G^{ij}_{k\ell} & = & \textrm{tr} ( (a^\dagger_i a_j - \gamma^i_j) (a^\dagger_{\ell} a_k - \gamma^{\ell}_k)  D ) \succeq 0,
\end{eqnarray}
but for a fixed particle state, these $G$-matrices share the same eigenvalues, since $\gamma^i_j$ can be replaced with
$\gamma^i_j \frac{1}{N} \sum_i a^\dagger_i a_i$ \cite{Erdahl74}. 
In Zhao {\it et al} \cite{Zhao04}, we can find explicitly formula of $T1$ and $T2$ conditions from Erdahl's survey paper \cite{Erdahl78}:
\begin{eqnarray}
 (T1)^{ijk}_{\ell mn} & = & \textrm{tr} ((a^\dagger_i a^\dagger_j a^\dagger_k a_n a_m a_{\ell}
+a_n a_m a_{\ell} a^\dagger_i a^\dagger_j a^\dagger_k ) D ) \succeq 0, \label{t1}\\ 
 (T2)^{ijk}_{\ell mn} & = & \textrm{tr} ( (a^\dagger_i a^\dagger_j a_k a^\dagger_n a_m a_{\ell}
+a^\dagger_n a_m a_{\ell} a^\dagger_i a^\dagger_j a_k ) D ) \succeq 0, \label{t2}
\end{eqnarray}
which are stronger conditions. 
An important property of these matrices is that $Q$, $G$, $T1$ and $T2$-matrices can be written {\it only} by linear combinations of the 2-RDM elements like following:
\begin{eqnarray}
 Q^{ij}_{k\ell} &=&(\delta^{i}_{k} \delta^{j}_{\ell} - \delta^{i}_{\ell} \delta^{j}_{k}) - (\delta^{i}_{k} \gamma^{j}_{\ell} + \delta^{j}_{\ell} \gamma^{i}_{k})+ (\delta^{i}_{\ell} \gamma^{j}_{k} + \delta^{j}_{k} \gamma^{i}_{\ell})- 2 \Gamma^{ij}_{k\ell},  \nonumber \\
 G^{ij}_{k\ell}&=&\delta^{j}_{\ell} \gamma^{i}_{k} - 2 \Gamma^{i \ell}_{k j}, \nonumber \\
(T1)^{i j k}_{\ell m n} & = & {\cal A} [i, j, k]  {\cal A} [\ell, m, n] (\frac{1}{6} \delta^{i}_{\ell} \delta^{j}_{m} \delta^{k}_{n} 
- \frac{1}{2}\delta^{i}_{\ell} \delta^{j}_{m} \gamma^{k}_{n}  + \frac{1}{2} \delta^{i}_{\ell} \Gamma^{j k}_{m n}), \nonumber \\
(T2)^{i j k}_{\ell m n} & = & {\cal A}[j, k] {\cal A}[m, n] (\frac{1}{2} \delta^{j}_{m}\delta^{k}_{n}\gamma^{i}_{\ell} + \frac{1}{2}\delta^{i}_{\ell} \Gamma^{m n}_{j k} -2\delta^{j}_{m}\Gamma^{i n}_{l k}), \nonumber
\end{eqnarray}
where $\cal A$ is the antisymmetrizer operator acting on an arbitrary function $f(i,j,k)$,
\[
{\cal A}[i,j,k]f(i,j,k) = f(i,j,k)-f(i,k,j)-f(j,i,k)+ f(j,k,i)+ f(k,i,j)- f(k,j,i).
\]
For $T1$ and $T2$'s cases, the 3-RDM terms cancel out. The $T2'$ condition replaces 
the $T2$ condition and is slightly strengthened by the addition of the one-particle operator \cite{Braams07,Mazziotti07}.

Other positive semidefinite type representability conditions are known such as the $B$ and $C$. However, they are implied by the $G$ condition \cite{Kummer1977}. 

We can extend these conditions to
positive semidefiniteness of higher order RDMs. These extensions seems to be known for a long time. Erdahl and Jin \cite{ErdahlJin} formulated the $p$-th order approximation to the $N$-particle density matrix in terms of the semidefiniteness conditions on the $p$-th order RDMs, which is an generalization of the $P$, $Q$, $G$, $T1$ and $T2^\prime$ conditions.

\subsection{The reduced-density-matrix method}
We call as the {\it reduced-density-matrix method}, the variational method 
having the 2-RDM (and the 1-RDM) as the basic variable(s)
restricted to some approximation ${\tilde {\cal E}}^N$ of the 
$N$-representability set ${\cal E}^N$. It can be formulated as follows:
\begin{equation}
{\tilde E}_{\rm min} 
      =  \underset{\Gamma \in {\cal \tilde E}^N}{\rm min} \{ \sum_{ij} v^{i}_{j} \gamma^{i}_{j} +  \sum_{ijk\ell} w^{ij}_{k\ell} \Gamma^{ij}_{k\ell} \}. 
\label{eq:approx}
\end{equation}

Among the possibilities, we usually consider the set obtained by imposing
some necessary conditions for the $N$-representability. 
The set $\tilde{\cal E}^N$ should satisfy the following properties.

\begin{itemize}
\item satisfies some necessary conditions of ensemble $N$-representability.
\item compact set, so that a linear functional (the Hamiltonian) has a minimum.
\item convex set, so that the solution would not be stuck into local minima.
\item stringent, so that resultant 2-RDM should be physically or chemically meaningful.
\item computationally feasible and/or efficient. 
\item completely general: since the form of the Hamiltonian is totally general, it is not only applicable to chemistry but also to physics.
\item {\it ab initio}: no empirical parameters. Currently very successful methods based on the density functional theories employ a lot of empirical parameters.
\end{itemize}

We can find new $N$-representability conditions from chemical or physical requirements satisfying the above properties by constructing a Hamiltonian and obtaining an upper bound to the ground state energy. Then we can add this as a new condition. These ``cuts'' may strengthen $\tilde {\cal E}^N$.

\begin{figure}
\begin{center}
\includegraphics[scale=0.4, bb=100 200 455 635]{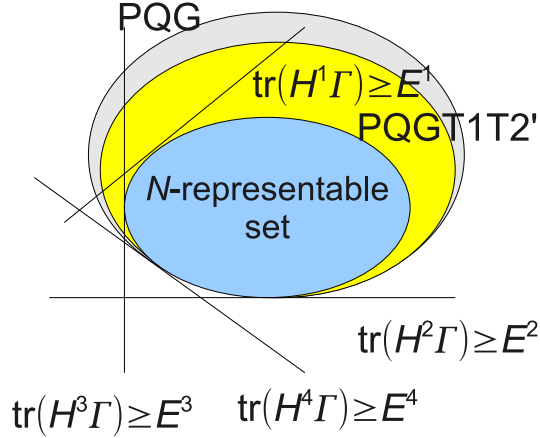}
\caption{Schematic representation of approximate $N$-representable sets; a better set shrinks.}
\label{fig:approxnrep}
\end{center}
\end{figure}

Trivial $N$-representability conditions with the $P$, $Q$, $G$, $T1$ and $T2^\prime$ conditions and every possible combination of the $P$, $Q$, $G$, $T1$, $T2$ and $T2^\prime$ conditions, satisfy the above criteria. These variational energies have the following property:
\begin{itemize}
\item If we add more necessary conditions, the calculated energy usually becomes better and would never become worse:
\[
 E_{PQ} \leq E_{PQG} \leq E_{PQGT1} \leq E_{PQGT1T2} \leq E_{PQGT1T2^\prime}
\leq E_{\rm full CI},
\]
\end{itemize}
where $E_{PQ}$ is the variational energy with the $P$ and $Q$ conditions, $E_{PQGT1T2}$ is the variational energy with the $P$, $Q$, $G$, $T1$ and $T2$ conditions, {\it etc.} (see Fig \ref{fig:approxnrep}). This property is totally {\it opposite} from traditional wavefunction approaches. Variational calculation using the wavefunction gives upper bounds, and approximation to the total energy becomes lower (better) when the variational space becomes larger .

Note that the obtained 2-RDM may not be unique, even when the original problem is non-degenerated and the energy is unique.


\subsection{Some interpretations on conditions}
We usually enforce only the necessary conditions on the trial 2-RDMs. Therefore, the RDM method gives lower bounds to the exact energy, an $N$-representable 1-RDM and a non-physical 2-RDMs. Thus it is important to realize the physical meaning of the necessary conditions employed in the calculations.
The following results may be useful to interpret results from actual calculations on molecules and atoms:
\begin{description}
\item {(a)} If a trial 2-RDM $\Gamma$ satisfies the $P$ and $Q$ conditions, the original 2-RDM is ensemble $N$-representable \cite{Coleman63}.  
\item {(b)} If a trial 2-RDM $\Gamma$ satisfies the $G$ condition, then 1-RDM from the original 2-RDM is ensemble $N$-representable \cite{Rosina69}. 
\item {(c)} If the Hamiltonian of a system is time-reversal invariant, and the number of particles $N$, is even, the necessary and sufficient condition that an approximate 1-RDM corresponding to a non-degenerate energy eigenstate be $N$-representable is that its natural spin-orbital occupation numbers be equal in pairs \cite{Smith66}. Moreover Coleman proved that the AGP (anti-symmetrized power) wavefunctions covers all such 1-RDMs \cite{ColemanYukalov}. Thus if
these conditions apply to the systems, we always obtain pure representable 1-RDMs with necessary $N$-representability conditions.
\item {(d)} The $G$ condition is related to the AGP type wavefunction, and gives the correct energy for the Hamiltonian for which the ground state can be written by the AGP function \cite{ErdahlRosina74}. Moreover, the AGP type wavefunction is closely related to the superconductivity \cite{Nakamura59}.
\item {(e)} The $G$ condition is exact at the high-correlation limit for the Hubbard model, since the two particle term of the model ($U\sum_{i}^L a^\dagger_{i\uparrow} a_{i\uparrow} a^\dagger_{i\downarrow} a_{i\downarrow}$, where $U>0$, $L$ is the number of sites, $i$ is the $i$-th site, and $\uparrow$ and $\downarrow$ denote up-spin and down-spin of the electrons, respectively) is a $G$-type Hamiltonian ($\sum_{ijk\ell} A^{ij}_{k\ell} a^\dagger_i a_j a^\dagger_{\ell} a_k$), which is bounded by zero \cite{Yamashita10}. 
\end{description}

\section{Formulation of the RDM method's Problem
as an Semidefinite Program and its Solution
by the Interior-Point Method}

\subsection{Semidefinite program}
\label{subsec:sdp}

Semidefinite program (SDP) has established as an important class of
problems in optimization since 1990's. 
It is known to have an elegant mathematical theory, and an
efficient algorithm called interior-point 
method which can solve it in polynomial-time complexity.
Refer for instance to \cite{Todd01} for a nice survey about SDPs.

Let $\C,\A_p \ (p=1,2,\ldots,u)$ be given block-diagonal real symmetric
matrices with prescribed block sizes, $\b\in\Real^u$ 
and $\c,\a_p\in\Real^s \ (p=1,2,\ldots,s)$ be given 
real vectors. 
We denote by $\Diag(\a)$ a diagonal matrix with the elements of 
the vector $\a$ on its diagonal.

An SDP is defined for instance by
\begin{equation}
\label{sdp:primal}
\left\{\begin{array}{ll}
\textrm{maximize} & \textrm{tr}(\C\X)+\textrm{tr}(\Diag(\c)\Diag(\x)) \\
\textrm{subject to} & \textrm{tr}(\A_p\X)+\textrm{tr}(\Diag(\a_p)\Diag(\x))=b_p, \quad (p=1,2,\ldots,u) \\
& \X\succeq\O, \ \x\in\Real^s, \\
\end{array}\right.
\end{equation}
where we refer it as the {\it primal SDP}.
The notation $\X\succeq \O$ means that $\X$ is symmetric positive semidefinite.
Then, we can define the {\it dual SDP} as
\begin{equation}
\label{sdp}
\left\{\begin{array}{ll}
\textrm{minimize} & \b^T\y \\
\textrm{subject to} & \S = \displaystyle \sum_{p=1}^u \A_py_p-\C\succeq \O, \\
& \displaystyle \sum_{p=1}^u\Diag(\a_p)y_p = \Diag(\c),\\
&\y\in\Real^u.\\
\end{array}\right.
\end{equation}
The variables for the primal SDP is $(\X,\x)$ while
for the dual SDP is $(\S,\y)$.
Under mild assumptions\footnote{to be more precise, we
need to eliminate some variables using the equalities
$\sum_{p=1}^u\Diag(\a_p)y_p = \Diag(\c)$, and assume the Slater's
condition, but we avoid to do it here to be cumbersome.} \cite{Todd01}, 
the solution of (\ref{sdp:primal}-\ref{sdp})
should satisfy
\begin{eqnarray}
\textrm{tr}(\A_p\X)+\textrm{tr}(\Diag(\a_p)\Diag(\x))=b_p, & (p=1,2,\ldots,u) \nonumber \\ 
\S = \displaystyle \sum_{p=1}^u \A_py_p-\C\succeq \O, & \nonumber \\
\X,\S\succeq \O, & \nonumber \\
\b^T\y - \textrm{tr}(\C\X)-\textrm{tr}(\Diag(\c)\Diag(\x)) = 0. & \label{gap}
\end{eqnarray}
These conditions are equivalent to earlier result by Erdahl \cite{Erdahl79}
and Bellman and Fan \cite{BF}.

The advantage of considering the variables in the primal and dual
SDPs simultaneously is that we can check the numerical correctness
of the approximate solution from the above relations.

\subsection{Formulation of the RDM method's problem as an SDP}

Hereafter, we assume that we have chosen $r$ spin orbitals 
from (\ref{eq:basis}), which is assumed to give a good approximation
for the desired wavefunction we seek.
There are plenty of these {\it bases} in quantum chemistry, and 
we actually use them on the numerical
experiments which follows. 
Also, notice that all definitions and notions of $N$-representability
and its conditions can be defined accordingly using this 
finite basis of CONS.

The RDM method's problem imposing some necessary
$N$-representability conditions 
such as the $P$ (\ref{p}), $Q$ (\ref{q}), $G$ (\ref{g}), $T1$ (\ref{t1}),
$T2$ (\ref{t2}) or $T2^\prime$ conditions,
is in fact an SDP. 
In order to make its formulation more clear, we
perform some linear transformations on the matrices involved in the
problem.
In (\ref{eq:approx}), the 1-RDM variable $\gam$, and the corresponding
Hamiltonian $\v$ have two indexes, which correspond to ordinary
matrices in linear algebra.
However, the other matrices
involved in the calculations have four or even six indices each. 
To convert from these notations convenient for quantum chemists to
the notation of elementary linear algebra, 
we need to map each pair $(i,j)$ or triple $(i,j,k)$ of indices 
to a composite
index on these matrices.
For instance, the 2-RDM element $\Gamma^{ij}_{k\ell}$ 
$(1\leq i < j \leq r; \ 1\leq k < \ell \leq r)$ will be mapped to
$\tilde{\Gamma}_{j-i+(2r-i)(i-1)/2,\ell-k+(2r-k)(k-1)/2}$, which
results in a symmetric matrix of size 
$r(r-1)/2 \times r(r-1)/2$.
We assume henceforth that all matrices are transformed to become
two-index matrices, and we keep the same notation as before for simplicity.
Furthermore, due to spin symmetry \cite{Zhao04}, all these matrices 
will reduce to
block-diagonal matrices of sizes specified in
Table~\ref{tab:size} \cite{Fukuda07,Fukuda07-2}
\footnote{There is an additional ${}_{r/2+1} C_2$ term in the column of $m$ of
Table~II [23], which is missing and corresponds to the size or the 1-RDM.}.

\begin{table}[!htbp]
\caption{Sizes of the SDP as a function of the number of spin orbitals $r$
for each necessary $N$-representability condition.}
{\tiny
\begin{tabular}{cl} \hline
$N$-repres. cond. & \multicolumn{1}{c}{size of block matrices} \\ \hline
$\gam\succeq\O$ & $r/2\times r/2$ (2 blocks) \\
$\I\succeq \gam$ & $r/2\times r/2$ (2 blocks) \\
$P$ condition & $\left(r/2\right)^2\times\left(r/2\right)^2$ (1 block), \
$\left(\begin{array}{@{\hskip.2\tabcolsep}c}r/2\\ 2\end{array}\right)\times
\left(\begin{array}{@{\hskip.2\tabcolsep}c}r/2\\ 2\end{array}\right)$ (2 blocks) \\
$Q$ condition & $\left(r/2\right)^2\times\left(r/2\right)^2$ (1 block), \ 
$\left(\begin{array}{@{\hskip.2\tabcolsep}c}r/2\\ 2\end{array}\right)\times
\left(\begin{array}{@{\hskip.2\tabcolsep}c}r/2\\ 2\end{array}\right)$ (2 blocks) \\
$G$ condition & $2\left(r/2\right)^2\times 2\left(r/2\right)^2$ (1 block), \
$\left(r/2\right)^2\times\left(r/2\right)^2$ (2 blocks) \\
$T1$ condition & $\frac{r}{2}
\left(\begin{array}{@{\hskip.2\tabcolsep}c}r/2\\2\end{array}\right)\times
\frac{r}{2}
\left(\begin{array}{@{\hskip.2\tabcolsep}c}r/2\\2\end{array}\right)$ (2 blocks), \
$\left(\begin{array}{@{\hskip.2\tabcolsep}c}r/2\\3\end{array}\right)\times
\left(\begin{array}{@{\hskip.2\tabcolsep}c}r/2\\3\end{array}\right)$ (2 blocks) \\
$T2$ condition & $\frac{r}{6}\left(\begin{array}{@{\hskip.2\tabcolsep}c}3r/2 \\ 2 \\
\end{array}\right)\times 
\frac{r}{6}\left(\begin{array}{@{\hskip.2\tabcolsep}c}3r/2 \\ 2 \\
\end{array}\right)$  (2 blocks), \
$\frac{r}{2}
\left(\begin{array}{@{\hskip.2\tabcolsep}c}r/2\\2\end{array}\right)\times
\frac{r}{2}
\left(\begin{array}{@{\hskip.2\tabcolsep}c}r/2\\2\end{array}\right)$ (2 blocks) \\ 
$T2^\prime$ condition & $\frac{r}{2}+\frac{r}{6}\left(\begin{array}{@{\hskip.2\tabcolsep}c}3r/2 \\ 2 \\
\end{array}\right)\times 
\frac{r}{2}+\frac{r}{6}\left(\begin{array}{@{\hskip.2\tabcolsep}c}3r/2 \\ 2 \\
\end{array}\right)$  (2 blocks), \
$\frac{r}{2}
\left(\begin{array}{@{\hskip.2\tabcolsep}c}r/2\\2\end{array}\right)\times
\frac{r}{2}
\left(\begin{array}{@{\hskip.2\tabcolsep}c}r/2\\2\end{array}\right)$ (2 blocks) \\ \hline
$u$ in (\ref{sdp}) & $\left(\begin{array}{@{\hskip.2\tabcolsep}c}
r^2/4+1\\ 2 \\ \end{array}\right)
+2\left(\begin{array}{@{\hskip.2\tabcolsep}c}r\left(r/2-1\right)/4+1 \\
2 \\ \end{array}\right)+\left(\begin{array}{c}\frac{r}{2}+1 \\ 2 \end{array}
\right)$ \\
$s$ in (\ref{sdp}) & $5 + 2\left(\begin{array}{@{\hskip.2\tabcolsep}c}r/2+1\\ 2 \\ \end{array}\right)$ \\ \hline
\multicolumn{2}{l}{here $\left(\begin{array}{c}a \\ b \end{array}\right)=
\frac{a!}{(b-a)!b!}$, for integers $a\geq b > 0.$}
\end{tabular}}\label{tab:size}
\end{table}

Now, let us define a linear transformation svec: $\SS^v\rightarrow
\Real^{v(v+1)/2}$ from the space of $v\times v$ symmetric matrices $\SS^v$.
For $\U\in\SS^v$,
\[
\textrm{svec}(\U) = (U_{11},\sqrt{2}U_{12},U_{22},\sqrt{2}U_{13},
\sqrt{2}U_{23},U_{33},\ldots,\sqrt{2}U_{1v},\ldots,U_{vv})^T.
\]
Then, defining 
$y=(\textrm{svec}(\gam)^T,\textrm{svev}(\Gam)^T)^T \in \Real^u$, 
$\b=(\textrm{svec}(\v)^T,\textrm{svev}(\w)^T)^T \in \Real^u$, 
and finding the suitable matrices $\C,\A_p \ (p=1,2,\ldots,u)$
and vectors $\c,\a_p \ (p=1,2,\ldots,u)$ for the corresponding 
necessary $N$-representability conditions of Table~\ref{tab:size}, for
instance, we can cast the problem
(\ref{eq:approx}) as (\ref{sdp}).

Although these transformations and formulations seem a little confusing, it
is in fact the formulation Garrod {\it et al.} arrived 35 years ago 
\cite{GF76}. 
Nakata {\it et al.} \cite{Nakata01,Nakata02,Nakata02-2} formulated the problem as the 
primal SDP (\ref{sdp:primal}) instead when they performed the first 
computation as an SDP. For a more detailed discussion about these
transformations and the formulations,
see for instance \cite{Zhao04,Fukuda07,Fukuda07-2}.

\subsection{Theoretical computational complexity of the primal-dual
interior-point method}

As it was previously mentioned, SDPs can be solved in 
polynomial-time using interior-point
methods \cite{Todd01}. In particular, employing the parallel code 
SDPARA \cite{SDPARA}, which is an implementation of the primal-dual
interior-point method, one can
theoretically expect that it will take 
$\OC(\sqrt{v_{\max}}\log\varepsilon^{-1})$ iteration with 
$\OC(u^2f^2/d+u^3/d+uv_{\max}^2+v_{\max}^3)$ floating-point operations 
per iteration.
Here $v_{\max}$ refers to the size of the largest block matrix in
$\A_p \ (p=1,2,\ldots,u)$, 
$f$ is the maximum number of nonzero elements in each of these matrices,
$d$ is the total number of available CPU cores in the parallel computer, and 
$\varepsilon$ is the accuracy which we can expect when we replace the
rhs of (\ref{gap}) ``=0'' by ``$\leq \varepsilon$'' (where $\varepsilon >0$).
In our case, $u=\OC(r^4)$, $v_{\max}=\OC(r^3)$, $f=\OC(1)$, and therefore,
the total theoretical floating-point operations is $\OC(r^{13.5}\log\varepsilon^{-1}/d)$
\cite{Fukuda07-2}.

\section{Some Historical Remarks}

Here we make an attempt to list some articles
related to our work in chronological order. 
However, it is far from being complete.

The definition of the RDM was explicitly spelled out by 
Husimi \cite{Husimi40} in 1940.
The dependence of the
energy on the 2-RDM (and 1-RDM) appeared in L\"owdin \cite{Lowdin55} 
and Mayer \cite{Mayer55} in 1955.
The necessary and sufficient conditions for an ensemble 1-RDM,
{\it i.e.}, $\0\preceq \gam \preceq \I$, tr$(\gam)=N$, were obtained by
Kuhn \cite{Kuhn60} in 1960 and Coleman \cite{Coleman63} in 1963.
In this latter article, the precise formulation of the 
$N$-representability problem,
the $P$ and $Q$ conditions (for the 2-RDM)
were also stated.
In the next year, the $G$ condition was proposed by
Garrod and Percus \cite{GarrodPercus64}.

The restriction of the $N$-representability problem only on the
diagonal elements of the 2-RDM, known as the {\it diagonal problem},
were investigated by Weinhold and Wilson \cite{WeinholdWilson67},
Davidson \cite{Davidson69}, McRae-Davidson \cite{MD72},
and Yoseloff \cite{Yoseloff74} since late 1960's. Other progresses
on this topic can be found in the survey \cite{Ayers07}. 

The first variational calculation on the 2-RDM imposing the necessary
$N$-representability conditions for the doubly ionized carbon C$^{++}$
$(N=4)$ were performed by Kijewski and co-authors 
since late 1960's \cite{Kijewski74} (see earlier reference therein).
Garrod and co-authors proposed several algorithms, some of them
which resemble modern optimization algorithms, and reported 
results for the beryllium ($N=4$) \cite{GMR75,RG75,GF76}.
In particular, Mihailovi\'{c} and Rosina applied it to nuclear
physics \cite{MR75}, but obtained large deviations to the full CI 
calculations if compared to
electronic systems.

In the 1978 survey paper of Erdahl \cite{Erdahl78}, we can find the
conditions which is knows as $T1$, $T2$ \cite{Zhao04}, and $T2^\prime$
conditions \cite{Braams07,Mazziotti07}. 
He also proposed algorithms based on the exact mathematical
characterization of solutions of the 
lower bound method (RDM method) in the next year \cite{Erdahl79}.

These were the golden ages for the RDM research, but somehow
faded away because it was realized soon that the underlying problem
is computationally difficult and poor results were obtained for nuclear systems.

A revival of the 2-RDM approach appeared since 1992 due to Valdemoro \cite{Valdemoro92}, Nakatsuji and Yasuda \cite{NakatsujiYasuda1997}, and Mazziotti \cite{Mazziotti98}. This approach is based on the density equation or
the contracted Schr\"odinger equation (CSE), which is equivalent to solve
the Schr\"odinger equation. Nakatsuji proved that if an $N$-representable 4-RDM satisfies the CSE, then the original $D$ satisfies the Schr\"odinger equation and vice versa \cite{Nakatsuji1976}.
Valdemoro, Nakatsuji-Yasuda and Mazziotti consider the 2-RDM as the basic variable. The CSE requires 1- to 4-RDMs, thus they reconstruct 3- and 4-RDMs using 1- and 2-RDMs and solve the CSE iteratively. In this approach, they assume the resultant 2-RDM is nearly $N$-representable because the reconstruction functional is physically relevant; they did not explicitly impose any $N$-representability conditions.
Their results are quite good, and can be compared to single and double CI for small atoms and molecules such as
$\rm Be$, $\rm Ne$, and $\rm CH_3F$. The absolute values of negative eigenvalues of the $P$, $Q$, and $G$-matrices were small. However, researchers payed little attention to this approach since non-convergence or divergence occur especially where the correlations are strong \cite{Ehara99}. There are difficulties in systematic refinements of the reconstruction functional even though some improvements are reported \cite{Yasuda99}, but there are even more miscellaneous problems \cite{Nooijen02}.

In 2001, Nakata {\it et al.} \cite{Nakata01} were the first to employ an 
optimization software to solve the RDM method's problem as
an SDP, and reported computational results imposing the
$P$, $Q$, and $G$ conditions on a series of small atoms and molecules.
The results were better than the SDCI calculations, and obtained correlation energies from 100\% to 120\%.
As mentioned in Section~\ref{subsec:sdp}, these
results have a numerical certificate of correctness, which
could not be obtained before the advent of interior-point methods.
Also, the numerical convergence does not depend on the initial
guess as it is common in Hartree-Fock, CCSD methods, {\it etc.} Moreover,
there exists a global minimum, whereas this is not guaranteed in the CSE approach
for instance.
In the next year, Nakata {\it et al.} \cite{Nakata02} showed results
for the dissociation limit for several molecules including triple bonded $\rm N_2$,
demonstrating numerically that the RDM method do not break down as the single reference
methods such as CCSD and perturbation methods. However, it was also shown that size consistency is slightly
deviated.

The inclusion of the Weinhold-Wilson inequalities \cite{WeinholdWilson67},
which was not satisfied only including the $P$, $Q$, and $G$ conditions \cite{Nakata02},
however showed little progress in the results \cite{Nakata02-2}.

In 2002, Mazziotti immediately reproduced Nakata {\it et al.}'s results and applied to diatomic molecules \cite{Mazziotti02}. The prolific research by Mazziotti and his colleagues in the following years \cite{Mazziotti07} corroborated with these results. 

In 2004, a breakthrough was done by Zhao {\it et al.} They included
additionally the $T1$ and $T2$ conditions,
which became very strong conditions for small molecules and atoms. They noted a ``spectacular increase in accuracy'' and results
were comparable to CCSD(T); typically the correlation energies for various atoms and molecules were between 100\% to 101\% \cite{Zhao04,Fukuda07,Nakata08}.

In the same year, Mazziotti announced the RRSDP method \cite{Mazziotti04}
in which he reformulates the SDP problem as an nonlinear and nonconvex problem and
applies a quasi-Newton method to solve it \cite{Burer03}. In 2006, Canc\`es {\it et al.} proposed and implemented the dual problem of
(\ref{eq:approx}) \cite{Cances2006}. 

Applications to the one-dimensional Hubbard model was done by Hammond {\it et al.} \cite{Hammond06}. They calculated the Hubbard models with $P$, $Q$, $G$, and $T2$ conditions up to 14 sites. The obtained error per site was $-0.0089$ for $L=14$'s case with $P$, $Q$, $G$, and $T2$ conditions for $U=8$, when correlation is strongest. Nakata {\it et al.} investigated the high correlation limit using multiple-precision arithmetic version of SDP solver, called the SDPA-GMP \cite{Nakata08}. At the high correlation limit, they reproduced the exact energy and proved that the $G$ condition is exact \cite{Yamashita10}. 

The size-consistency and size-extensivity are important properties when the size of the systems becomes bigger or larger. Nakata {\it et al.} found slight deviations \cite{Nakata02}, but Van Aggelen {\it et al.} showed very clear and systematic examples that the RDM method gives incorrect dissociation limit with fractional charges on the well-separated atoms of diatomic molecules with $P$, $Q$ and $G$ conditions. For the $\rm CO$'s case, 
the Mulliken populations were 5.98 and 8.00 at the dissociation limit.
Even adding $T1$ and $T2$ conditions they did not fix the problem \cite{Aggelen2009,Aggelen2010}. Nakata and Yasuda 
investigated numerically that size-extensivity is also slightly violated by calculating 32 non-interacting
$\rm CH_4$  and $\rm N_2$ \cite{Nakata2009}. The inextensive contributions to energies are $3\times 10^{-4}$
 and $3\times 10^{-3}$ a.u. using the STO-6G basis set, respectively.
Later, Verstichel {\it et al.} also proposed a method to ``cure'' this pathological behaviour of the RDM method, however quite demanding \cite{Verstichel2010}. Currently solutions to size-consistency is not practical.

\section{Numerical Results for the RDM Method}

Here, we give numerical results we
obtained so far for the RDM method imposing some necessary
$N$-representability conditions. Some of them are completely new.

\subsection{New numerical results for larger systems}

We present here some numerical results for the largest systems
solved so far by our group.

The SDPs obtained by the RDM method imposing
the $P$, $Q$, $G$ or $P$, $Q$, $G$, $T1$, $T2^\prime$ conditions were
solved using the parallel code SDPARA~7.3.2 \cite{SDPARA,Fujitsu2010}.
The calculations were performed at 
the Kyoto University's T2K supercomputer using
128 nodes, were each node has 4 CPUs (quad-core AMD Opteron 8356 2.3GHz)
and 32GB of memory, giving a total of  2048 cores; and
at a self-made computer cluster using 16 nodes, were each node has 
2 CPUs (quad-core Intel Xeon 5460 3.16GHz) and 48GB of memory,
giving a total of 128 cores.

Table~\ref{tab:new1} shows the results for five molecules were
$r=28,30$ or $36$ spin orbitals were used \cite{Fujitsu2010}.
The full CI and SDCI (singly and doubly substituted configuration interaction)
calculations were performed using the package Gamess
\cite{Gamess},
while CCSD(T) (coupled cluster singles and doubles with perturbational treatment
of triples) and Hartree-Fock calculations were obtained by Gaussian98
\cite{Gaussian98}.
The entries, excepting the full CI, give the ground state energy 
differences to the full CI. 
The RDM method always gives a energy lower than full CI, while
SDCI and Hartree-Fock give higher. CCSD(T) usually results in higher
energy, but not necessarily. Units are in Hartree.
The acceptable accuracy in quantum chemistry is 1kcal/mol which
corresponds to approximately 0.0016 Hartree.
Also, the correlation energy $\varepsilon_{\textrm{corr}}$
is an important measure in quantum 
chemistry. It is defined as a percentage relative to the Hartree-Fock
(0\%) and full CI (100\%):
\[
\varepsilon_{\textrm{corr}}=
\frac{|E-E_{\textrm{HF}}|}{E_{\textrm{HF}}-E_{\textrm{FCI}}} \times 100,
\]
where $E$ the energy calculated by the RDM method, CCSD(T) or SDCI.

\begin{table}[!htbp]
\caption{Ground state energies (in differences from that of full CI) calculated by the
RDM method imposing the $P$, $Q$, $G$, $T1$, $T2^\prime$ conditions from
SDPARA~7.3.2, and those obtained by CCSD(T), SDCI, and Hartree-Fock from 
Gamess and Gaussian98. The last column shows the full CI energies. The 
energy units are in Hartree ($=4.3598\times 10^{-18}$J).
The correlation energies (0\% for Hartree-Fock and 100\% for full CI) in
percentage are also shown in the second row.}
{\tiny \begin{tabular}{c@{\hskip.5\tabcolsep}c@{\hskip.5\tabcolsep}c@{\hskip.5\tabcolsep}c@{\hskip.5\tabcolsep}c@{\hskip.5\tabcolsep}c@{\hskip.5\tabcolsep}c@{\hskip.5\tabcolsep}c@{\hskip.5\tabcolsep}c@{\hskip.5\tabcolsep}c@{\hskip.5\tabcolsep}c} \hline
system & state & basis & $r$ & $N(N_\alpha)$ & $2S+1$ & $\Delta E_{\textit{PQGT1T2$^\prime$}}$ & $\Delta E_{\textit{CCSD(T)}}$ & $\Delta E_{\textit{SDCI}}$ & $\Delta E_{\textit{HF}}$ & $E_{\textit{FCI}}$ \\ \hline
NH$_2^-$ & ${}^1A_1$ & double-$\zeta$ & 28 & 10 (5) & 1 & $-$0.000~6 & $+$0.000~63 & $+$0.008~74 & $+$0.141~98 & $-$55.624~71 \\ 
& & & & & & 100.4 & 99.55 & 93.84 & 0 & 100 \\
CH$_2$ & ${}^1A_1$ & double-$\zeta$& 28 & 8 (4)  & 1 & $-$0.000~4 & $+$0.000~59 & $+$0.005~80 & $+$0.100~67 & $-$38.962~24 \\
& & & & & & 100.4 & 99.42 & 94.24 & 0 & 100 \\
NH$_3$ & ${}^1A_1$ & valence double-$\zeta$ & 30 & 10 (5) & 1 & $-$0.000~5 & $+$0.000~49 & $+$0.007~46 & $+$0.128~75 & $-$56.304~89 \\
& & & & & & 100.4 & 99.62 & 94.45 & 0 & 100 \\
CH$_3$ & ${}^2A_2^{''}$& valence double-$\zeta$ & 30 & 9 (5)  & 2 & $-$0.000~3 & $+$0.000~31 & $+$0.004~01 & $+$0.094~54 & $-$39.644~14 \\ 
& & & & & & 100.3 & 99.67 & 95.75 & 0 & 100 \\
C$_2$ & ${}^1\Sigma_g^+$ & valence double-$\zeta$ & 36 & 12 (6)  & 1  & $-$0.003~5 & $+$0.000~39 & $+$0.055~98 & $+$0.285~66 & $-$75.642~11 \\   
& & & & & & 101.2 & 99.86 & 80.41 & 0 & 100 \\ \hline
\end{tabular}}\label{tab:new1}
\end{table}

From Table~\ref{tab:new1}, we can conclude that the RDM method imposing the
$P,Q,G,T1,T2^\prime$ conditions, gives equally better energies than
CCSD(T) in absolute value. C$_2$ molecule is exceptional, however, it is
known to be a difficult system in quantum chemistry.

Table~\ref{tab:new2} shows the same result for the O$_2^+$ molecule were $r=40$
spin orbitals were used.
The full CI calculation was not possible for this case due to the size limit on
the computer, and therefore, we restrict to show only the ground
state energy corresponding for each entry.

\begin{table}[!htbp]
\caption{Ground state energies calculated by the
RDM method imposing the $P$, $Q$, $G$ conditions from
SDPARA~7.3.2, and those obtained by CCSD(T), SDCI, and Hartree-Fock from 
Gamess and Gaussian98. The energy units are in Hartree ($=4.3598\times 10^{-18}$J).}
{\tiny
\begin{tabular}{c@{\hskip.5\tabcolsep}c@{\hskip.5\tabcolsep}c@{\hskip.5\tabcolsep}c@{\hskip.5\tabcolsep}c@{\hskip.5\tabcolsep}c@{\hskip.5\tabcolsep}c@{\hskip.5\tabcolsep}c@{\hskip.5\tabcolsep}c@{\hskip.5\tabcolsep}c} \hline
system & state & basis & $r$ & $N(N_\alpha)$ & $2S+1$ & $E_{\textit{PQG}}$ & $E_{\textit{CCSD(T)}}$ & $E_{\textit{SDCI}}$ & $E_{\textit{HF}}$ \\ \hline
O$_2^+$ & $^2\Pi_g$& double-$\zeta$ & 40 &  15 (8) &  2 & $-$149.450~2 & $-$149.385~95 & $-$149.360~26 & $-$149.091~83 \\ \hline
\end{tabular}}\label{tab:new2}
\end{table}

Table~\ref{time} shows the typical sizes of the problem when
formulated as an SDP (see Section~\ref{subsec:sdp}), and
the computational time to solve them by SDPARA~7.3.2 at the T2K supercomputer
or at the computer cluster.

In the previous work \cite{Nakata08}, we only could solve the RDM method's
problem with $P,Q,G,T1,T2^\prime$ conditions up to $r=28$ spin orbitals.
Here, we give results up to $r=36$. This achievement was possible
due to a major update in the parallel code SDPARA~7.3.2 \cite{SDPARA,Fujitsu2010}.
It became faster, and now it can take advantage of multi-core 
(multi-thread) computation in addition to the ordinary
MPI (message passing interface) computation.

\begin{table}[!htbp]
\caption{Size of SDPs obtained by the RDM method imposing some $N$-representability
conditions shown at Tables~\ref{tab:new1} and~\ref{tab:new2},
and their computational time when solved at the 
T2K supercomputer or at the computer cluster (c.c.).}
{\begin{tabular}{cclrrrrr} \hline
system & $r$ & $N$-repres. cond. & $u$ & $v_{\max}$ & time (s) & system & CPU cores \\ \hline
NH$_2^-$ & 28 & $P,Q,G,T1,T2^\prime$ & 27,888 & 4,032 & 27,949 & T2K & 2048 \\
CH$_2$ & 28 &$P,Q,G,T1,T2^\prime$ & 27,888 & 4,032 & 26,656 & T2K & 2048 \\
NH$_3$ & 30 & $P,Q,G,T1,T2^\prime$& 36,795 & 4,965 & 72,026 & T2K & 2048 \\
CH$_3$ & 30 & $P,Q,G,T1,T2^\prime$& 36,795 & 4,965 & 68,593 & T2K & 2048 \\
C$_2$ & 36 & $P,Q,G,T1,T2^\prime$ & 76,554 & 8,604 & 1,554,675 & c.c. & 128 \\
O$_2^+$  & 40 &$P,Q,G$ & 116,910 & 800 & 5,943 & T2K & 2048 \\ \hline
\end{tabular}}\label{time}
\end{table}

For other physical properties such as the dipole moments, refer to
\cite{Nakata08}.

\subsection{Summary of the numerical experiments}

We present a graphical summary of the data obtained in our
previous work \cite{Nakata08} with the addition of new ones presented in
the previous section.

\begin{figure}[!htbp]
\begin{center}
\includegraphics[scale=0.65]{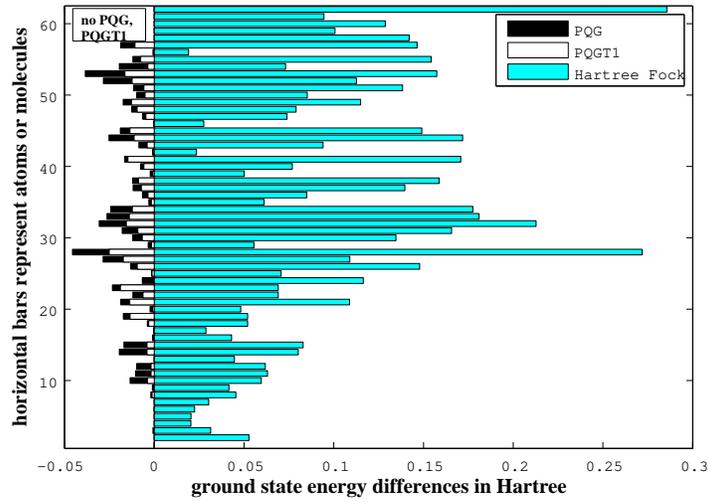}
\caption{Ground state energies (in differences from that of full CI) calculated by the RDM method imposing the $P,Q,G$, or $P,Q,G,T1$ conditions,
and those obtained by Hartree-Fock for the 57 atomic and molecular systems
of [45] and the 5 of Table~\ref{tab:new1}.} 
\label{fig:dif1}
\end{center}
\end{figure}

The ground state energy differences to full CI of the RDM method
imposing the $P,Q,G$ or $P,Q,G,T1$, and of Hartree-Fock for the
57 atomic or molecular systems \cite{Nakata08} and those 5 shown at 
Table~\ref{tab:new1} are plotted in Figure~\ref{fig:dif1}.
Each horizontal bar corresponds to a system and they are ordered
accordingly to the order it appears in the tables. That is, the
lowest one corresponds to 
the Lithium atom with $r=10$ spin orbitals \cite{Nakata08}, 
while the upper 5 corresponds to the data of Table~\ref{tab:new1}
(notice that we do not have the values for $P,Q,G$ and
$P,Q,G,T1$ entries for this case).

Apparently, it does not seems to exist an correlation between
the Hartree-Fock and the RDM method's results. However, we clearly notice
that the imposing the $P,Q,G$ conditions, we can obtain results
much better than the Hartree-Fock's ones.

\begin{figure}[!htbp]
\begin{center}
\includegraphics[scale=0.65]{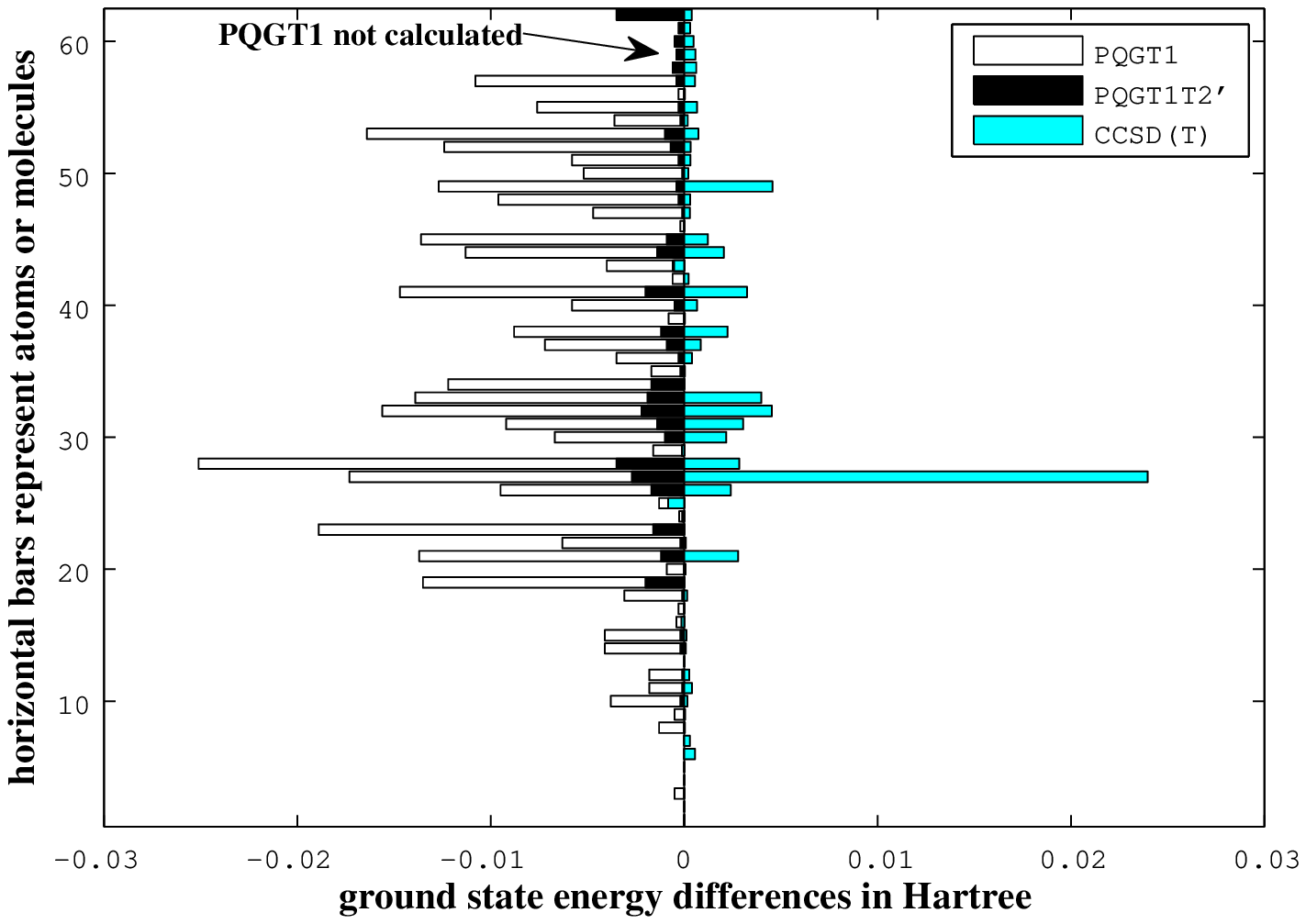}
\caption{Ground state energies (in differences from that of full CI) calculated by the RDM method imposing the $P,Q,G,T1$, or $P,Q,G,T1,T2^\prime$ conditions,
and those obtained by CCSD(T) for the 57 atomic and molecular systems
of [45] and the 5 of Table~\ref{tab:new1}.}
\label{fig:dif2}
\end{center}
\end{figure}

Figure~\ref{fig:dif2} shows the ground state energy 
differences to full CI of the RDM method
imposing the $P,Q,G,T1$ (same value as Figure~\ref{fig:dif1})
or $P,Q,G,T1,T2^\prime$, and of CCSD(T) for the
57 atomic or molecular systems \cite{Nakata08} and those 5 shown at
Table~\ref{tab:new1}
(notice that we do not have the values for $P,Q,G,T1$ entries for this case).

If you compare the RDM method with $P,Q,G,T1,T2^\prime$ conditions and
the CCSD(T) values, they seems equally good.
However, in some case CCSD(T) can fails to converge. There are
4 case in \cite{Nakata08} which are replaced by zero in Figure~\ref{fig:dif2}.
The largest deviation of 0.00279 Hartree to full CI for CCSD(T) is for
the Oxygen atom at the $^1D$ state \cite{Nakata08}.

\section{Concluding Remarks}
In this chapter, we showed the outline of the reduced-density-matrix method with applications to atomic and molecular fermionic systems. Some feature of this method are: (i) it is an {\it ab initio} method, which is rigorously the same as the Schr\"odinger equation for the ground state; (ii) the number of variables is always four, regardless of the size of the system; (iii) from the sparsity of the first- and second- order reduced density matrices the existence of a linear scaling method is apparent. The major obstacle for this method is the fundamentally difficulty of obtaining the complete $N$-representability conditions for the 2-RDM. However we know fairly good approximated (necessary) condition like $P$, $Q$, $G$, $T1$ and $T2^\prime$ conditions which reproduces comparable ground state energies to CCSD(T), which is considered the
golden standard method in quantum chemistry. 
The considered problem becomes a semidefinite programming problem, which minimizes a linear functional keeping the eigenvalues of matrices non-negative. 
In this chapter, we presented new results; $\rm NH_2^-$, $\rm CH_2$, $\rm NH_3$, $\rm C_2$, $\rm CH_3$, and $\rm O_2^+$ using a supercomputer with a highly efficient semidefinite programming solver, SDPARA. The semidefinite programming problems for $\rm NH_3$, $\rm C_2$, $\rm CH_3$, and $\rm O_2^+$ are the largest problems solved so far in the standard formulation. The correlation energies using $P$, $Q$, $G$, $T1$, $T2^\prime$ for $\rm NH_2^-$, $\rm CH_2$, $\rm NH_3$ was $100.4\%$, for $\rm CH_3$ was $100.3\%$, and for $\rm C_2$ was $101.2\%$, respectively. For $\rm O_2^+$ we used the double-$\zeta$ basis, and attained $132\%$ of correlation energy since the open-shell systems are difficult and due to large space, we only employ the $P$, $Q$ and $G$ conditions. 

We would like to close this chapter saying that the RDM method is a promising method for quantum chemistry or condensed matter physics. Developing the RDM method is important and fundamental to chemistry and physics.

\section*{Acknowledgments}
The large-scale supercomputer computations for this research have been
supported by the Collaborative Research Program for Large-Scale Computation
of ACCMS and IIMC, Kyoto University.
M.~N. was supported by the Special Postdoctoral 
Researchers' Program of RIKEN, and is partially supported by Grant-in-Aid 
for Scientific Research (B) 21300017.
M.~F. is very thankful for the invitation to the program
``Complex Quantum Systems'' held at the IMS-NUS, specially to the organizers
Heinz Siedentop and Matthias Christandl. 
He also enjoyed the discussion with Prof. Robert Erdahl.
M.~F. is partially supported by Grant-in-Aid for 
Young Scientists (B) 21700008.

\end{document}